\documentclass{PoS}

\title{Higgs--Cosmology Interplay}

\ShortTitle{Higgs--Cosmology Interplay}

\author{\speaker{Oleg Lebedev}\\
        Department of Physics, University of Helsinki, Gustaf H\"allstr\"omin katu 2a,
Helsinki, Finland \\
        E-mail: \email{oleg.lebedev@helsinki.fi}}

\abstract{I discuss the role of the Higgs boson as a probe of dark matter and inflation.\
          }

\FullConference{The European Physical Society Conference on High Energy Physics\\
		5-12 July, 2017\\
		Venice}

\begin{document}

\section{Introduction}

The discovery of the Higgs boson  at the Large Hadron Collider (LHC) in 2012 marked a new era in particle physics. Not only was it the last missing piece of the Standard Model (SM), it is also the only fundamental spin--0 particle known to date.  Its existence is crucial for quantum consistency of the SM and many of its measured properties conform to the theoretical predictions. Yet, the Higgs sector remains one of the least explored areas, leaving ample room for ``new physics'' effects.

The Higgs boson, being the only known scalar particle, enjoys a special status since it can interact directly with the ``hidden world'', that is, states which have no SM charges and thus  invisible to standard probes. In particular, dark matter (DM) and the field responsible for cosmological inflation (``inflaton'') can both belong to this category.
Therefore, the Higgs offers a unique probe of   the dark side of the Universe, whose existence has firmly been  established through  cosmological/astrophysical observations.

\section{The Higgs and dark matter}

A unique feature of the Higgs field is that $H^\dagger H$ is the only gauge and Lorentz invariant dim-2 operator that can be composed of the SM fields. 
\begin{figure}[!]
\begin{center}
\includegraphics[height=80mm]{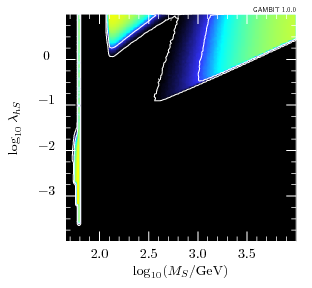}
\caption{Parameter space of the scalar singlet DM model. $M_S$ is the DM mass and $\lambda_{hs}$ is the Higgs portal coupling.
The area within the white contour is allowed at $2\sigma$,   while the lighter--colored   areas are 
favored statistically. Figure credit: GAMBIT collaboration \cite{Athron:2017kgt}. }
\label{f1}
\end{center}
\end{figure}
Therefore, the Higgs can couple to the hidden sector scalars (SM singlets)  $S$ at the renormalizable level,
\begin{equation}
-{\cal L} = \lambda_{hs} H^\dagger H S^\dagger S \;, \label{1}
\end{equation}
where we have assumed that the linear term in $S$ is forbidden by symmetry.
In the simplest scenario, a real scalar $S$  itself plays the role of dark matter \cite{Silveira:1985rk}. Indeed,
it is stable by virtue of  (\ref{1}) and couples only weakly to the SM fields. In particular, the DM scattering off nuclei is suppressed by the small Higgs--nucleon coupling.
A recent parameter space analysis performed by the GAMBIT collaboration  \cite{Athron:2017kgt}  shows that, 
despite its  simplicity and rigidity, 
this   model is viable. Their study includes various direct and indirect DM detection bounds, the relic DM abundance constraint as well as that from the Higgs decay at the LHC.   
The allowed parameter space in terms of $\lambda_{hs}$
and the DM mass $M_S$ is shown in Fig.~\ref{f1}.  The ligher shades in this plot 
indicate regions  favored by a combination of the constraints, while the area within the white contour is allowed at $2\sigma$.  One observes that DM can be as light as about 300 GeV at weak coupling (and away from the Higgs resonance region). However, the scan favors the multi--TeV range of $M_S$. This is a consequence of the ever--improving direct detection constraints, e.g. from XENON1T,  which push the DM mass to higher values. One may interpret this tendency as a crisis of the WIMP paradigm, however such a conclusion would be highly model dependent. 
If one adds at least one more degree of freedom to the dark sector, the direct detection rate can effectively be decorrelated from the DM annihilation cross section.

Consider the simplest generalization of the above model, whereby one complexifies the singlet $S$ (see, e.g. \cite{McDonald:1993ex}). It is natural to impose on the potential  a global U(1): $S \rightarrow e^{i\alpha} S$. To avoid the appearance of the Goldstone boson, however, one also needs a soft breaking mass term $S^2$. Adding these ingredients together, one finds  \cite{Gross:2017dan}
   \begin{eqnarray}
&&V=V_0 + V_{\rm soft} \;,\nonumber\\
&&V_0= 
-\frac{ \mu_H^2}{2} \, |H|^2 
- \frac{\mu_S^2}{2} \, |S|^2 
+\frac{\lambda_h}{2} |H|^4 
+\lambda_{hs} |H|^2 |S|^2 
+ \frac{\lambda_s}{2} |S|^4 \;, \nonumber\\
&&V_{\rm soft} =
- \frac{\mu_S^{\prime 2}}{4} \, S^2 + \textrm{h.c.} \label{pot}
\end{eqnarray}
\begin{figure}[!]
\begin{center}
\includegraphics[height=70mm]{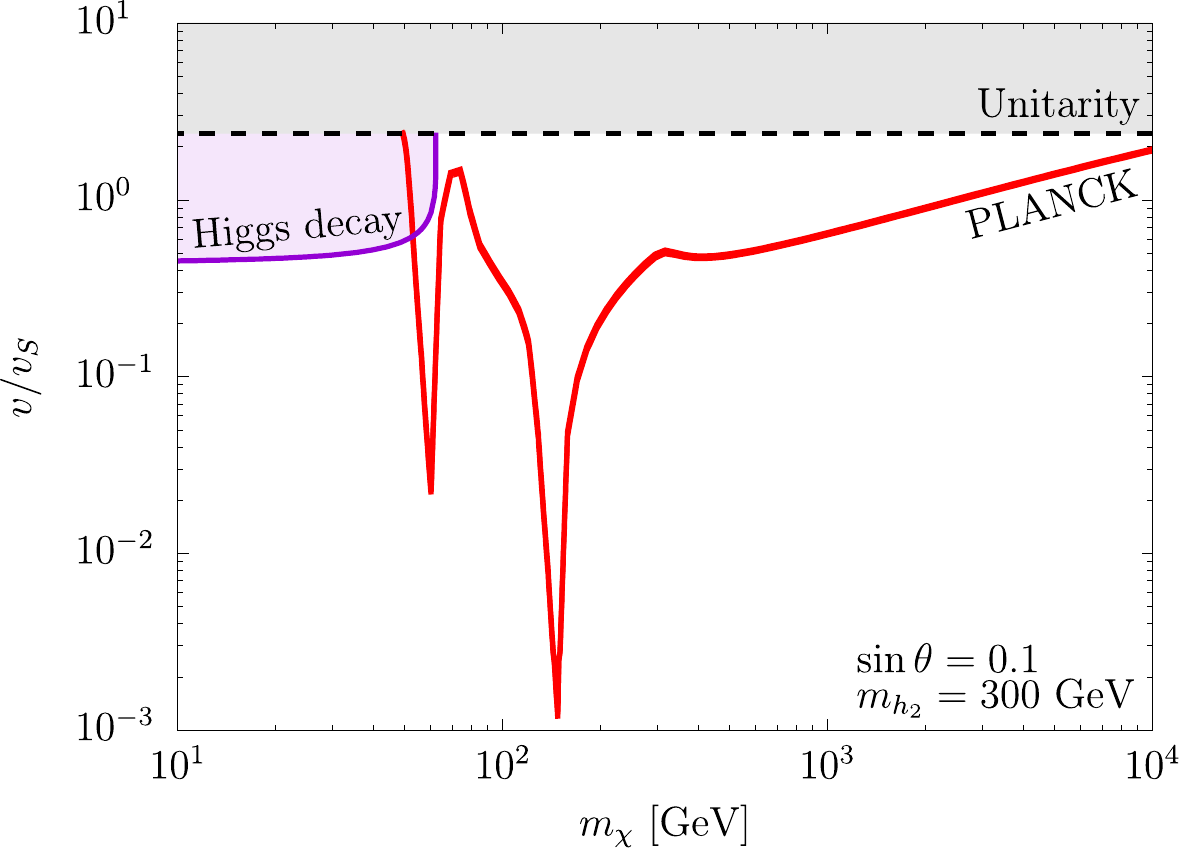}
\caption{Parameter space of the complex scalar model (``pseudo--Goldstone'' DM) \cite{Gross:2017dan}.  Here $m_\chi$ is the DM mass, $v_S=\sqrt{2} \langle S\rangle$, and $v=246$ GeV.  The red band leads to the correct DM relic abundance, while the purple and grey areas are excluded by the Higgs decay and perturbative unitarity, respectively.   }
\label{f2}
\end{center}
\end{figure}
After rotating away the phase of $\mu_S^{\prime 2}$, all the parameters of the model become real leading to the CP symmetry $S \rightarrow S^*$, which is also preserved by the vacuum. This makes Im$S$ stable and a natural DM candidate. A remarkable consequence of this simple set--up is that the direct DM detection amplitude vanishes at tree level and zero momentum transfer \cite{Gross:2017dan},
\begin{equation}
 \mathcal{A}_{dd}(t) \propto \sin \theta \cos \theta \left(\frac{m_2^2}{t-m_2^2}-\frac{m_1^2}{t-m_1^2}\right)   \simeq 0 \;
\end{equation}
 where $m_{1,2}$ are the masses of the CP even scalars of the model, $\theta$ is the Higgs--singlet mixing angle and $t$ is the Mandelstam variable,
 $t/m_{1,2}^2 \ll 1$. The cancellation occurs for $any$ parameter choice and is only spoiled by loop effects.  This can be traced back to the {\it (pseudo-) Goldstone} nature of dark matter: it is equivalent to the angular component of $S=\rho e^{i\phi}$, $\phi$, whose interactions vanish at zero momentum transfer. Introduction of the mass term $S^2$ does not affect the relevant vertex $\phi \phi \rho$ which vanishes for $\phi$ on shell and zero momentum of $\rho$.

The allowed parameter space of the model is shown in Fig.~\ref{f2}. 
One finds that the direct detection constraints are very weak and superseded by the 
perturbative unitarity bounds. The cancellations in the direct detection amplitude do not apply to the annihilation processes due to a large momentum transfer.
As a result,  our dark matter is a standard WIMP  whose mass is allowed to be anywhere between 60 GeV and 10 TeV by the above constraints. This illustrates that even very simple Higgs portal models can naturally  evade the XENON1T bounds.

In the above models, dark matter owes its stability  to a global U(1) or Z$_2$. 
One may argue that gauge symmetry is better motivated and dark matter models based on it would be more attractive. Vector Higgs portal DM offers a simple framework where dark matter stability results from gauge symmetry \cite{Hambye:2008bq,Lebedev:2011iq,Gross:2015cwa}. The basic idea is that Lie groups possess
inner and outer automorphisms which can play the role of dark matter stabilizers.  For example, in the simplest case of U(1) gauge symmetry the corresponding (outer) automorphism is complex conjugation of the group elements which corresponds to charge conjugation in physics terms. 
Consider a U(1) gauge theory with a single charged scalar $\phi$ (with charge +1/2) \cite{Lebedev:2011iq,Gross:2015cwa},
 \begin{equation}
 {\cal L_{\rm hidden}}= -{1\over 4} F_{\mu\nu} F^{ \mu\nu} + (D_\mu \phi)^\dagger D^\mu \phi -V(\phi) \;.
 \end{equation}
At the minimum of the scalar potential $V(\phi)$,
 $\phi$ develops a VEV, $\langle \phi \rangle= 1/\sqrt{2}~ \tilde v$. The imaginary part of $\phi$ gets $ $eaten by the $ $gauge field which
 now $ $acquires the mass $m_A= \tilde g \tilde v /2$, with $\tilde g$ being the gauge coupling. Decomposing 
  $\phi=1/\sqrt{2}~ (\rho +\tilde v)$ , we get the following gauge--scalar interactions:
  \begin{eqnarray}
&&  \Delta {\cal L}_{\rm s-g}= {\tilde g^2\over 4} \tilde v \rho \; A_\mu A^{ \mu} +
 {\tilde g^2\over 8}  \rho^2 \; A_\mu A^{\mu}  \;. 
 \end{eqnarray}
The system $ $possesses the  $Z_2$ symmetry 
\begin{equation}
A_\mu \rightarrow - A_\mu  ~,
\end{equation} 
which corresponds to charge conjugation of the original fields, 
$\phi \rightarrow \phi^*$ and $A_\mu \rightarrow - A_\mu$. Adding the Higgs portal term 
$\lambda_{h \phi} \vert H \vert^2 \vert \phi \vert^2$ leads to the $h-\rho$ mixing without affecting the symmetry. As a result, $A_\mu $ can play the  role of WIMP dark matter since the annihilation channels into the SM particles become available. One finds that the model   satisfies all of the constraints in substantial regions of parameter space \cite{Lebedev:2011iq,Gross:2015cwa}. 

This idea can be generalized  to non--Abelian models as well. One finds that if the hidden gauge symmetry is broken to nothing using the minimal ``hidden Higgs'' content, there is a subset of massive gauge fields which are stable \cite{Gross:2015cwa}. The stabilizing symmetry  depends on the group and further details, e.g. whether the CP symmetry is broken by the scalar potential.
For instance, in the SU(3) case with unbroken CP, the symmetry is $U(1)\times Z_2$ \cite{Arcadi:2016kmk}.

\begin{figure}[!]
\begin{center}
\includegraphics[height=80mm]{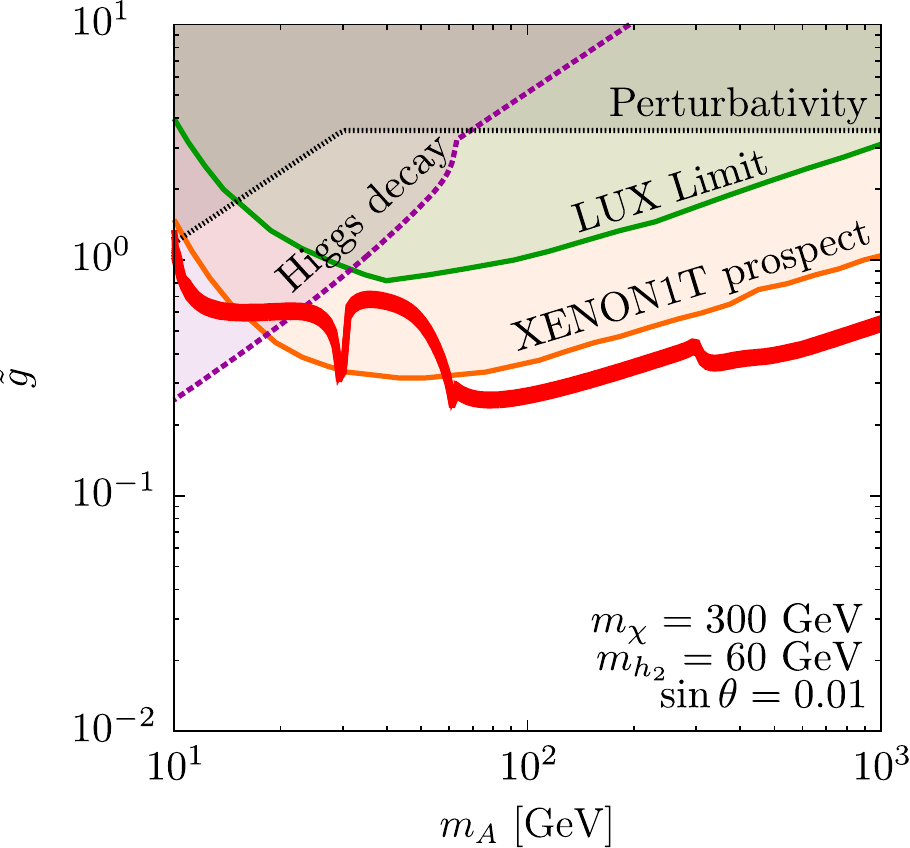}
\caption{Parameter space of the SU(3) vector   DM model  \cite{Arcadi:2016qoz}. $m_A$ is DM mass and $\tilde g$ 
is the hidden sector gauge coupling.
The red band leads to the correct relic DM abundance, while the areas above the purple and green lines are excluded by the Higgs decay and direct DM detection data, respectively.
    }
\label{f3}
\end{center}
\end{figure}

 If the hidden sector has more than one degree of freedom, an interesting option of ``secluded DM'' \cite{Pospelov:2007mp} becomes available. In this case, the main annihilation channel for dark matter could be pair production of the   hidden sector states which subsequently decay into SM particles. For instance, in the above U(1) example, one may have
 \cite{Arcadi:2016qoz}
 \begin{equation}
 A_\mu + A_\mu \rightarrow h_2 +h_2 \rightarrow {\rm SM ~fields} ~,
 \end{equation}    
 where $h_2=  h \; \sin\theta  +   \rho \; \cos\theta$ is assumed to be lighter than dark matter  $A_\mu$.  This process is not suppressed by small $\theta$ (which only affects the lifetime of $h_2$), while the direct detection rate vanishes as $\theta \rightarrow 0$.
This decorrelates the  annihilation cross section from the direct detection bounds thus allowing for a wide range of the WIMP masses. Fig.~\ref{f3} shows an example of ``secluded DM'' in the SU(3) vector DM model with $\sin\theta=0.01$. In this case, DM can annihilate both into unstable hidden gauge fields and pairs of $h_2$. The direct detection bounds 
are loose   allowing  for DM mass from tens of GeV to multi--TeV  \cite{Arcadi:2016qoz}.

We conclude that despite increasing pressure from direct DM detection experiments, Higgs portal dark matter is viable and offers an attractive WIMP framework with the DM mass ranging from tens of GeV to about 10 TeV. Some of the light DM regions can be probed at the LHC, e.g. via monojet searches with missing energy  \cite{Kim:2015hda}.

\section{The Higgs and inflation}

The Higgs portal coupling (\ref{1}) has an important impact on the Higgs dynamics in the Early Universe. The current data favor metastability of the electroweak vacuum 
\cite{Buttazzo:2013uya,Bezrukov:2012sa,Alekhin:2012py} which poses 2 puzzles \cite{Lebedev:2012sy}: why the Universe has chosen an energetically disfavored state and why it stayed there during inflation. In particular, even if the initial conditions of the Higgs field were fine--tuned such that it took on a value close to the origin, field fluctuations during inflation tend to destabilize it  \cite{Espinosa:2007qp,Espinosa:2015qea,East:2016anr}.  This behavior, however, is affected by the presence of the Higgs portal interactions. Already a  tiny coupling to an extra scalar can stabilize the Higgs potential
\cite{Lebedev:2012zw,EliasMiro:2012ay}, while the minimal option would be 
to modify the Higgs potential during inflation  only by including  the Higgs--inflaton interaction  \cite{Lebedev:2012sy}.  A similar effect is achieved via a non-minimal Higgs coupling   to gravity \cite{Espinosa:2007qp,Herranen:2014cua}.

On general grounds one expects the presence of the following interaction terms \cite{Ema:2017loe},
\begin{eqnarray}
&&  -{\cal L}_{h\phi } = \lambda_{h \phi} H^\dagger H \phi^2 + \sigma H^\dagger H \phi \,,\\
&& -{\cal L}_{hR } = \xi H^\dagger H \hat R \;, \label{L}
 \end{eqnarray}
where $\phi$ is the inflaton, $\hat R$ is the Ricci scalar  and $\xi$ is a constant.
The trilinear term $ H^\dagger H \phi$ arises in realistic inflation models since a linear inflaton coupling is required for  reheating \cite{Gross:2015bea}. All of these couplings are generated radiatively, even if absent at tree level. 

During inflation, the above interactions create an effective mass  for the Higgs field.
Assuming the simplest inflation model with $V_{\rm infl}= m^2 \phi^2/2$ and omitting the smaller $\sigma$--term, the 
effective mass is positive in the Einstein frame if $ \lambda_{h \phi} + 2m^2 \xi>0 $. 
If this combination is greater than about $10^{-10}$ and the initial inflaton value is large enough, the Higgs is so heavy that it 
 evolves quickly to zero, while the inflaton undergoes a slow roll \cite{Lebedev:2012sy}. The upper bound on 
  $ \lambda_{h \phi} + 2m^2 \xi $ is set by the requirement that the Higgs--inflaton coupling not spoil the flatness of the inflaton potential. When these conditions are satisfied, the above mentioned cosmological problems are resolved: during inflation, the Higgs potential is dominated by the positive quadratic term which pushes the field to smaller values while also suppressing its quantum fluctuations.

  \begin{figure}[t]
\begin{center}
\includegraphics[width=6.8 cm]{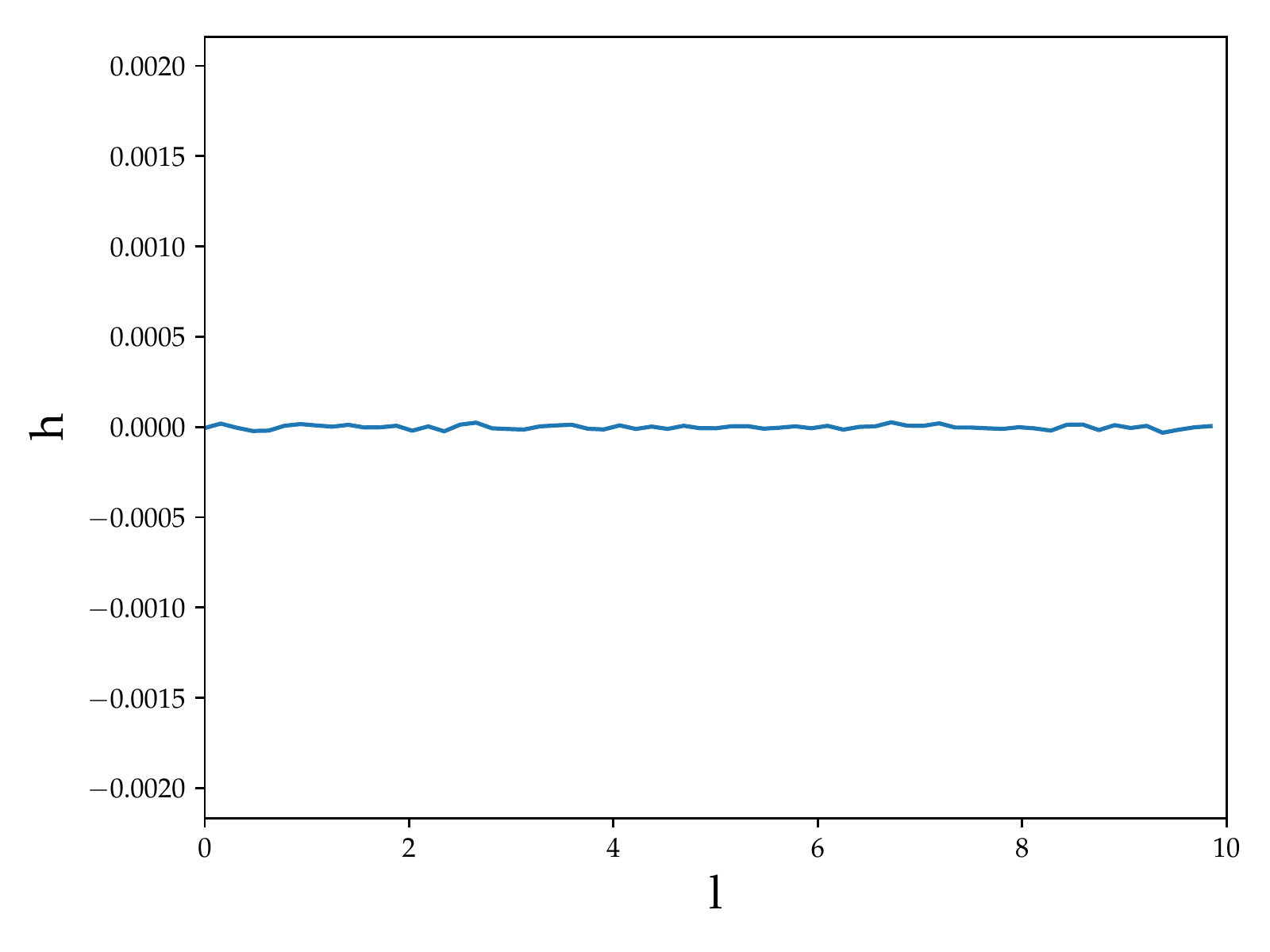}
\hspace{0.3cm}
\includegraphics[width=6.8 cm]{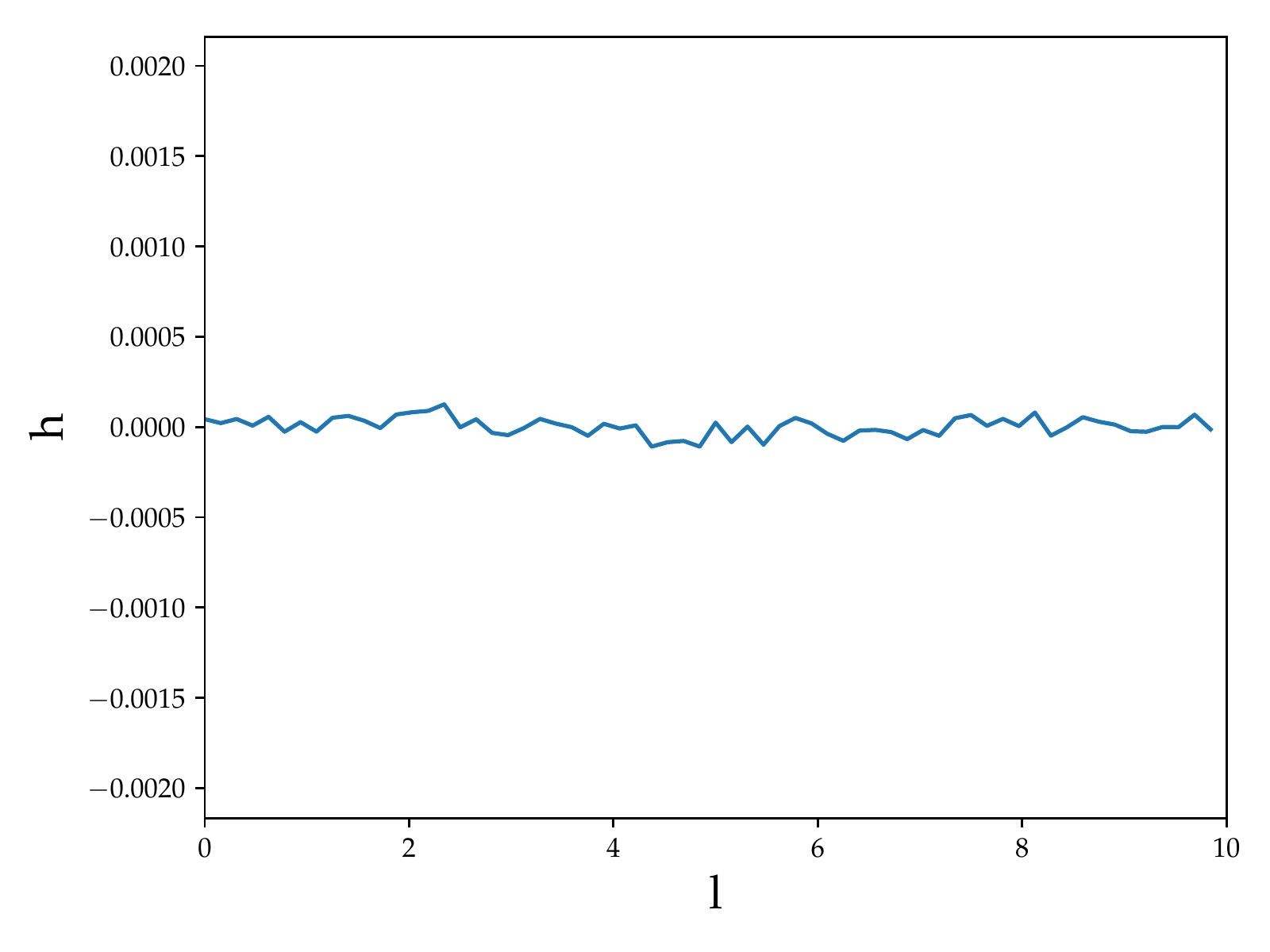}\\
\includegraphics[width=6.8 cm]{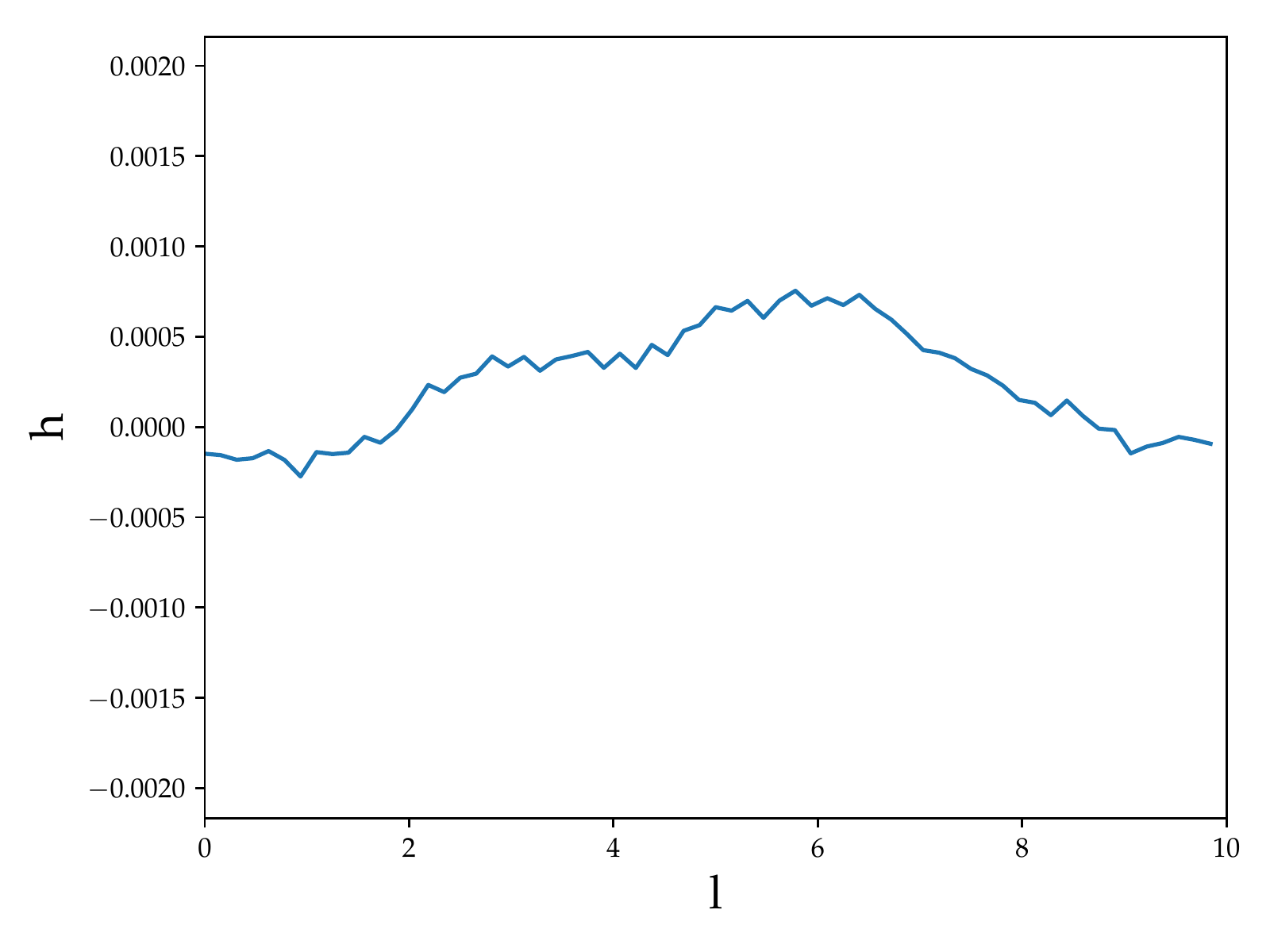}
\hspace{0.3cm}
\includegraphics[width=6.8 cm]{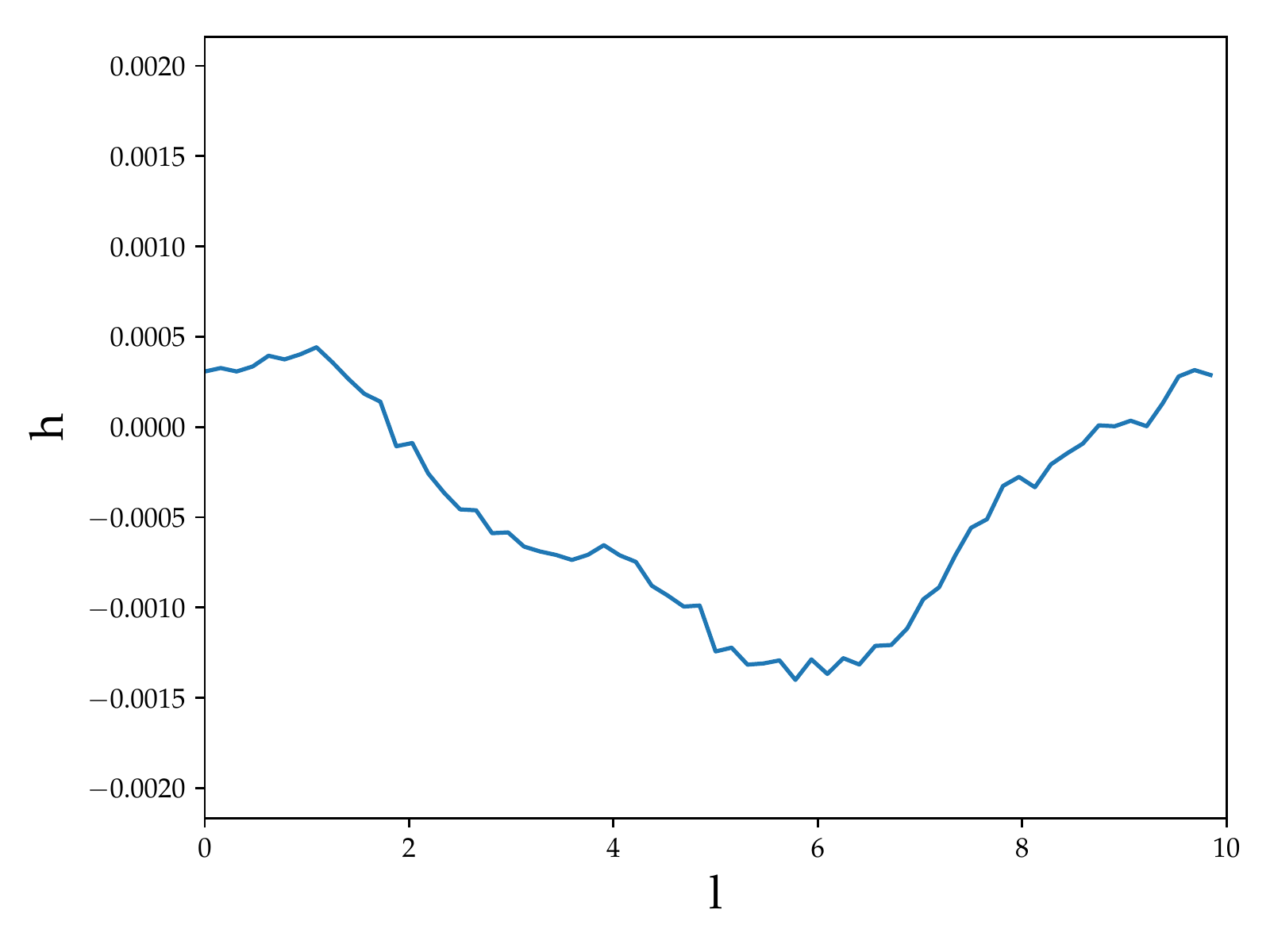}
\end{center}
\caption{Growth of Higgs fluctuations during preheating. $h$ is the Higgs field value at the lattice site with position $l$ in units of the inverse inflaton mass, $m^{-1}$.  The frames display the field profiles at (left to right, top to bottom) $mt=0, 15, 27, 30$,
obtained with LATTICEEASY \cite{Felder:2000hq}.
 }
\label{Fluc}
\end{figure}

Although the Higgs--inflaton/gravity couplings can stabilize the Higgs during inflation, they may have a destabilizing effect immediately after inflation, i.e. during the inflaton oscillation/preheating epoch. In particular, the resonant production of Higgs quanta characteristic of this period can be so efficient that the resulting field fluctuations 
(Fig.~4) drive the Higgs field towards the catastrophic vacuum \cite{Mind}. 
The particle production occurs due to parametric \cite{Kofman:1994rk} and tachyonic  \cite{Felder:2000hj} resonances induced by 
the Higgs--inflaton/gravity couplings. The resulting model constraints for the case of  a single coupling  in Eq.~\ref{L} have been studied in Refs.~\cite{Herranen:2015ima} and \cite{Ema:2016kpf,Kohri:2016wof}. The case of multiple couplings was considered in 
Refs.~\cite{Enqvist:2016mqj} and \cite{Ema:2017loe}.

  Let us discuss the main features of Higgs production during preheating following 
  Ref.~\cite{Ema:2017loe}. The analysis is conveniently performed in terms of 
  the rescaled (spacial)  Fourier $k$--modes of the Higgs field, $X_{k}\equiv a^{3/2}h_{k}$, where  $a$ is the scale factor. Defining the time variable $z=mt/2$, one finds
  \begin{equation}
\frac{\mathrm{d}^{2}X_{k}}{\mathrm{d}z^{2}}+\left[A_{k}\left(z\right)+2p\left(z\right)\cos 2z
+2q\left(z\right)\cos 4z  + {\delta m^2 (z) \over m^2}\right]X_{k}
\simeq0.\label{EoM-preh}
\end{equation}
 Here $A_k, p,q$ are combinations of the couplings and the inflaton field amplitude, e.g. $p$ and $q$ are effective Higgs mass terms induced by the trilinear and  quartic/$\xi$ couplings, respectively.  
 In this equation, we have resorted to  the Hartree approximation $h^3 \rightarrow 3h \langle h^2 \rangle$ such that  the Higgs self--interaction is replaced by an effective mass term $\propto \delta m^2$. In this case, the equations for different $k$--modes decouple. (This approximation is not employed in lattice simulations.)

Eq.~\ref{EoM-preh} is similar to the Whittaker--Hill  equation apart from the (slow)
 time--dependence of the coefficients $A_k,\delta m^2 , p,q$. The resonant behavior of the solutions 
 is  stipulated by the cosine terms. When the coefficients lie in an instability band, the solutions grow exponentially in time. However, since $A_k, p,q$  (slowly) descrease, the system exits the band and enters a stable regime where $X_{k}$   oscillate. When $A_k, p,q$
decrease further, $X_k$ experience exponential growth again. This happens until the system reaches the last stability band where it remains ($p,q \lesssim 1 $).  While at the early stages $ \delta m^2$ is negligible, towards the end of the resonance it grows in magnitude 
 and, due to its negative sign at large field values, leads to further amplification of $X_{k}$.

The growth of the Higgs  Fourier modes $X_{k}$ can be interpreted as a growth in the occupation numbers $n_k$ associated with the corresponding frequency $\omega_k$. The total Higgs variance is then calculated according to 
\begin{equation}
 \langle h^2\rangle \simeq \int \frac{d^3 k }{(2\pi a)^3}\frac{n_k}{\omega_k}\,.
 \end{equation}
If $\langle h^2\rangle^{1/2}$ exceeds the distance to the potential barrier separating the electroweak vacuum from the catastrophic one, $h_{\rm crit} \sim \sqrt{2(\lambda_{h \phi} + 2m^2 \xi)/|\lambda_h|} |\phi|$, the system gets destabilized. We find 
that in the absence of $\xi$, stability during preaheating requires 
\begin{equation}
\lambda_{h \phi} \lesssim 3 \times 10 ^{-8 } ~~,~~\vert \sigma\vert \lesssim 10^8 \; {\rm GeV}\;,
\end{equation}
assuming the SM critical scale of $10^{10}$ GeV and quadratic inflaton potential. The presence of the non-minimal Higgs coupling to gravity relaxes the constraint on 
$\lambda_{h \phi}$ as shown in Fig.~5. In this case, the parametric resonance can be suppressed due to an interplay between $\lambda_{h \phi}$ and $\xi$, allowing for 
$\lambda_{h \phi}$ above $10^{-6}$ and $|\xi|$ up to $10^4$.

\begin{figure}[!]
\begin{center}
\includegraphics[height=70mm]{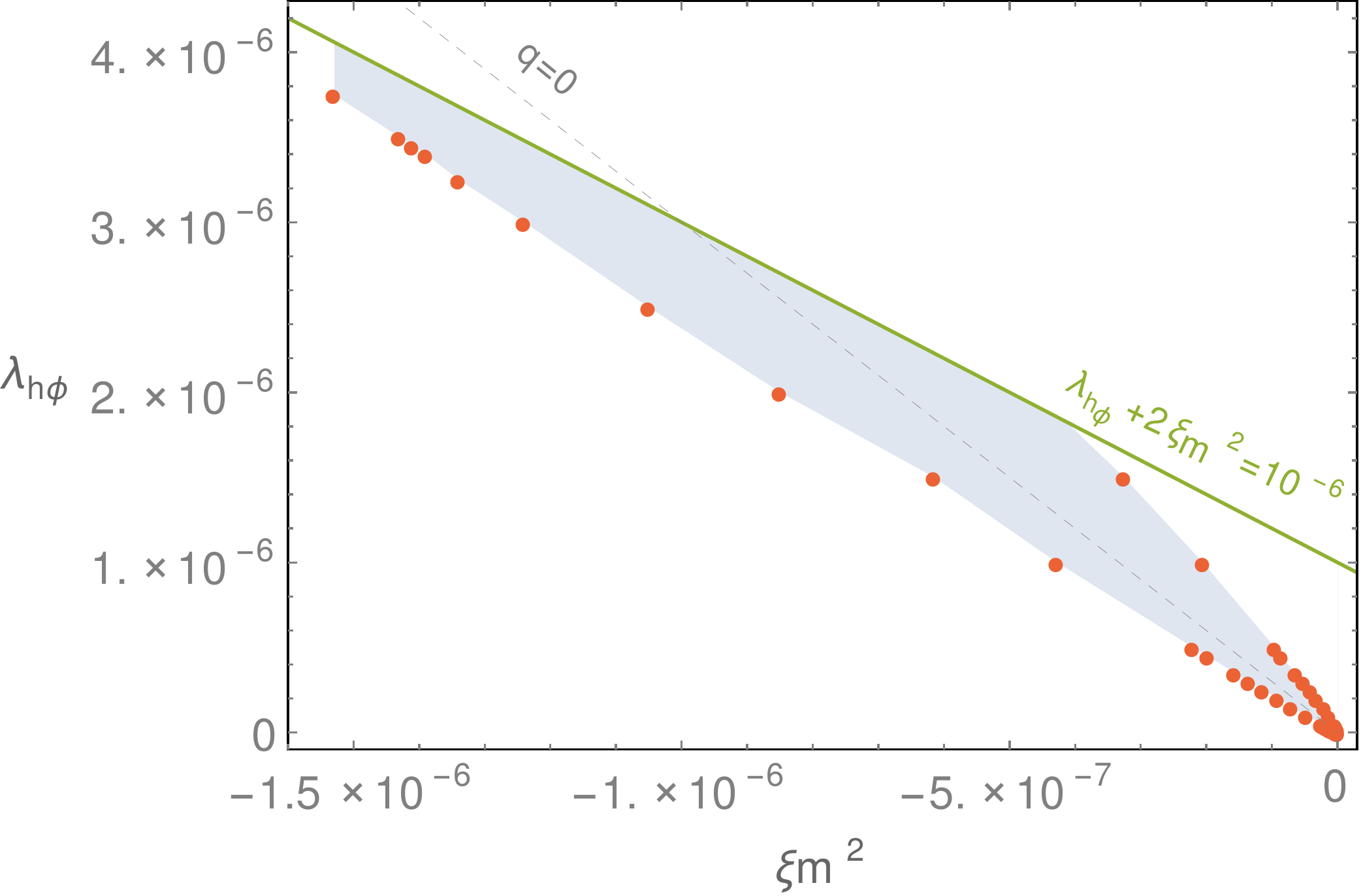}
\caption{Coulings consistent with Higgs potential stability during inflation and preheating
(shaded) \cite{Ema:2017loe}. Here $m^2 \simeq 10^{-10}$ in Planck units and the bisector of the shaded region ($q=0$ line) corresponds to complete suppression of the parametric resonance. }
\label{bounds}
\end{center}
\end{figure}

The above constraints inherit some degree of model dependence which is to be quantified in future work. In particular, an essential role in these considerations can be played by the Higgs--inflaton mixing which is allowed to be as large as 0.3 by the low energy data.
 Further work is required to map out all the viable possibilities.

To summarize this section,   Higgs--inflaton and Higgs--gravity interactions play a crucial role in understanding the Early Universe Higgs dynamics, especially in light of apparent metastability of the electroweak vacuum.\footnote{This statement is of course sensitive to the precise value of the top quark mass, which may change in the future. If the electroweak vacuum is stable, a very interesting option of Higgs inflation becomes available \cite{Bezrukov:2007ep}. }

\section{Conclusion}

The Higgs field offers a unique probe of the hidden sector which may host dark matter and an inflaton. The framework of Higgs portal dark matter offers a number of simple viable options, including models with ``pseudo--Goldstone''  or ``secluded'' dark matter. On the other hand, the Higgs--inflaton interaction is essential for reconstructing the cosmological evolution of the Higgs field culminating in the electroweak vacuum state.

\end{document}